%% file: scalapack_wn.tex
\documentclass[10pt]{article}

\usepackage[utf8]{inputenc}
\usepackage{fullpage}
\usepackage{amsmath, amssymb, amsthm}
\usepackage{tikz,pgfplots}
\usepackage[space,noadjust]{cite}
\usepackage{tabularx}
\usepackage{multirow}
\usepackage{rotating}
\usepackage{graphicx}
\usepackage{grffile}
\usepackage{caption,subcaption}
\usepackage[colorlinks=true]{hyperref}
\usepackage[frozencache]{minted}
\usepackage{algorithmic}
\usepackage{algorithm}

\usepackage{todonotes}
\newcommand{\tr}{T}

\newcommand{\flame}{\textsc{FLAME }}
\newcommand{\slicot}{\textsc{SLICOT }}
\newcommand{\blas}{\textsc{BLAS }}
\newcommand{\clapack}{\textsc{CLAPACK }}
\newcommand{\cuda}{\textsc{CUDA }}
\newcommand{\cusolver}{\textsc{cuSOLVER }}
\newcommand{\scalapack}{\textsc{ScaLAPACK }}
\newcommand{\lapack}{\textsc{LAPACK }}
\newcommand{\linpack}{\textsc{LINPACK }}
\newcommand{\matlab}{\textsc{MATLAB }}

\newcommand{\tolthreez}{\textsc{TOL3Z }}

\newcommand{\roff}{{\boldsymbol\varepsilon}}

\definecolor{mintedbg}{rgb}{0.95,0.95,0.95}

\newtheorem{example}{Example}
\newtheorem{remark}{Remark}

\hypersetup{%
	pdftitle = {A note on the ScaLAPACK routines for computing the QR-factorization with column pivoting}, %
	pdfauthor = {Zvonimir Bujanovic, Zlatko Drmac}, %
}

\title{New robust ScaLAPACK routine for computing the QR factorization with column pivoting}%
\author{Zvonimir Bujanović\thanks{University of Zagreb, Department of Mathematics, Zagreb, Croatia; {\tt zbujanov@math.hr}; {\tt drmac@math.hr}}
\and Zlatko Drmač\footnotemark[1]}

\begin{document}
\maketitle

\begin{abstract}
In this note we describe two modifications of the \scalapack subroutines \texttt{PxGEQPF} for computing the QR factorization with the Businger-Golub column pivoting. First, we resolve a subtle numerical instability in the same way as we have done it for the LAPACK subroutines \texttt{xGEQPF, xGEQP3}  in 2006. [LAPACK Working Note 176 (2006); ACM Trans. Math. Softw. 2008]. The problem originates in the first release of \linpack in the 1970's : due to severe cancellations in the down-dating of partial column norms, the pivoting procedure may be in the dark completely about the true norms of the pivot column candidates. This may cause miss-pivoting, and as a result loss of the important rank revealing structure of the computed triangular factor, with severe consequences on  other solvers that rely on the rank revealing pivoting. The instability is so subtle that e.g. inserting a \texttt{WRITE} statement or changing the process topology can drastically change the result. Secondly, we also correct a programming error in the complex subroutines  \texttt{PCGEQPF, PZGEQPF}, which also causes wrong pivoting because of erroneous use of \texttt{PSCNRM2, PDZNRM2} for the explicit norm computation. 
%\todo{Will update Abstract.}
\end{abstract}

\input{01_intro}
\input{02_numex}
\input{03_correction}
\input{04_conclusion}
%%%%%%%%%%%%%%%%%%%%%%%%%%%%%%%%%%%%%%%%%%%%%%%%%%%%%%%%%%%%%%%%%%%%%%%%%
\bibliographystyle{plain}
\bibliography{scalapack_wn.bib}
\end{document}

%% file: 01_intro.tex
%!TEX root=scalapack_wn.tex
\section{Introduction}
In our 2006.~paper \cite{DrmBuj08} we revealed a  subtle numerical instability in the \lapack \cite{LAPACK} implementations \texttt{xGEQPF, xGEQP3} of the QR factorization with the Businger-Golub column pivoting \cite{bus-gol-65}. 
Recall, if $A\in\mathbb{C}^{m\times n}$, then the pivoted QR factorization reads
\begin{equation}\label{eq:QRCP}
A \Pi = Q \begin{pmatrix} R\cr 0\end{pmatrix}
\end{equation}
\begin{equation}\label{QR-structure}
|R_{11}|\geq |R_{22}|\geq\cdots \geq |R_{nn}| ; \;\;
|R_{ii}|\geq \sqrt{\sum_{k=i}^j |R_{kj}|^2}=\|R(i:j,j)\|,\;\;\mbox{for all}\;\;1\leq {i}\leq {j} \leq n .
\end{equation}
The structure (\ref{QR-structure}) is the key for a rank revealing property of the factorization; the pivoting strives to maximize $|R_{ii}|$ at each step, thus globally trying to maximize the volume (absolute value of the determinant) of $R$, which is an important mechanism for the strong rank revealing property of the factorization. Further, the strong diagonal dominance of $R$  enhances the accuracy of the computed factorization and the stability of the backward substitutions, e.g., in solving the least squares problems \cite{DrmBuj08}.  Due to the special structure (\ref{QR-structure}), the matrix $R_r = \mathrm{diag}(1/\|R(i,:)\|)_{i=1}^n R$ is well conditioned independent of the condition number of $A$, which is the key ingredient in a Jacobi type SVD method \cite{drm-99-Jacobi}, \cite{dgesvd-99}, \cite{drm-ves-VW-1, drm-ves-VW-2}.  The factorization (\ref{eq:QRCP}) is also at the kernel of the QDEIM method \cite{QDEIM}, \cite{WQDEIM} which is a useful tool in nonlinear model order reduction and optimal sensor placement. In finite precision, the inequalities (\ref{QR-structure}) may hold up to a small roundoff, which is acceptable \cite{DrmBuj08}. 
%% QDEIM ..... 

Although (\ref{QR-structure}) was specified in the definitions of \texttt{xGEQPF} and  \texttt{xGEQP3}, we were able to construct examples for which  it failed dramatically in both subroutines -- the $|R_{ii}|$'s where not monotonically decreasing and did not dominate the remaining sub-columns, and the numerical rank of $A$ was severely underestimated. What made the problem more intriguing was the fact that changing compiler options or strategically placing \texttt{WRITE} statement (to display the value of a particular variable) could restore or destroy (\ref{QR-structure}), thus strikingly changing the numerical rank of $A$ and, e.g., the solution of the least squares problem. The numerical robustness and reproducibility in scientific computing are important and  nontrivial issues, and if a strategically placed \texttt{WRITE} statement, or different optimization level, can change the output dramatically as a result of ill-conditioning, then we ought to take this problem seriously.
Furthermore, the debuggability of such a code using modern debugging tools is questionable, as the debugging mode is only  a simulation of an actual run, with possibly entirely different behaviour. This also calls for serious rethinking of the compiler and debugger design for numerical computations. For an in depth discussion and more examples related to this issue we refer to the lecture notes by Kahan \cite{kahan-BASCD-2008}, \cite{kahan-STANF50-2008}, \cite{kahan-NEEDEBUG-2011},  \cite{kahan-Boulder-2012}.

 Just glancing through the dependency tree of \lapack reveals that the list of affected solvers in \lapack includes  \texttt{xGELSX} and \texttt{xGELSY} (for solving the least squares problem $\|Ax-b\|\rightarrow\min$),  \texttt{xGGSVP} (deprecated), \texttt{xGGSVP3, xGGSVD3} (GSVD of matrix pairs $(A,B)$),  \texttt{xGEJSV} (Jacobi SVD\footnote{It was the development of this routine \cite{drm-ves-VW-1, drm-ves-VW-2} that exposed the problem.}). Recently, it was shown in \cite{drm-xgesvdq} that (\ref{QR-structure}) was the key ingredient to make the QR SVD as accurate as the Jacobi SVD.

In \cite{kahan-Boulder-2012}, Kahan argues that the incidence of misleadingly inaccurate computed results is higher than generally believed, and as one kind of evidence he discusses \emph{"Revelation, after long use, that a widely trusted program produces, for otherwise innocuous input data, results significantly more inaccurate than previously believed."} He continues with  \emph{"The longest instance I know about was exposed by Zlatko Drma\v{c} \& Zvonimir Bujanovi\'{c} [2008, 2010] in a program used heavily by \linpack, \lapack, \matlab and numerous others since 1965 to estimate ranks of matrices."}

The problem originated from the initial description of the algorithm in the 1965.~paper \cite{bus-gol-65} and  has spread across software libraries  from the first implementation of (\ref{eq:QRCP}, \ref{QR-structure}) in the \linpack subroutine \texttt{xqrdc} in the 1970's.  Our solution to this problem, described in detail in \cite{DrmBuj08}, was included in the \lapack 3.1.0 release on November 12, 2006.
For software packages that use the \lapack as a computing engine, the problem has been automatically resolved simply by linking the new version of the \lapack library. This includes \clapack and, e.g., the \cusolver in the \cuda Toolkit. Some other packages have used the source code of the \lapack subroutines in their own subroutines, and the problem persists (and keeps spreading further) unless an explicit action is taken to implement the correction described in \cite{DrmBuj08}.   For example,  we have done this for the SLICOT library; as described in \cite{SLWN-2010-1}, $60$ out of $470$ subroutines in the SLICOT library (2010.~release) subroutines had been found susceptible to this problem. Our update is also included in \texttt{NoFLA\_HQRRP\_WY\_blk\_var4}, \texttt{NoFLA\_QRP\_downdate\_partial\_norms} in the \flame package HQRRP  \cite{2015arXiv151202671M}.

Unfortunately, some implementations of the column pivoted QR factorization, such as the \texttt{xGEQPX} in  \cite{bischof-q-orti-RRQR-1998-TOMS782} and \texttt{PxGEQPF} in \scalapack, still contain this hidden instability. If \texttt{PxGEQPF} is used as a model for other parallel HPC implementations (e.g., a \cuda version for CPU+GPU clusters), then  the focus is on reducing the communication and the numerical part will be inherited and the latent culprit will keep on messing up the diagonal dominance (\ref{QR-structure}). This is precisely how the critical part of the code was transplanted from \texttt{xqrdc} into \texttt{xGEQPF, xGEQP3, PxGEQPF, xGEQPX} and many others. An additional peculiarity of the problem (in addition to the previously mentioned sensitivity of the pivoting even to inserting a \texttt{WRITE} statement) is that various instances of miss-pivoting can be obtained (with the same input matrix) simply by changing the topology of the processes' grid; adding more processors may change the result dramatically. This is indubitably unacceptable behaviour for a scientific computing software.

The time is ripe for finally removing this problem from \scalapack and other relevant libraries.
In this note, we provide a more robust version of  \texttt{PxGEQPF}, which contains modification analogous to the changes we introduced in
\texttt{xGEQPF, xGEQP3} \cite{DrmBuj08, drmac-bujanovic-LWN-176-2006}. In addition, we correct a programming error in the complex subroutines \texttt{PSCNRM2, PDZNRM2}, which had similar damaging effects to the structure of the triangular factor.

The rest of this note is organized as follows. In \S \ref{S=2}, we illustrate the problem using a numerical example, and in \S \ref{SS=2.1} we briefly review the mathematical details of the source of the error. In \S \ref{S=3} we display the critical parts of the source code of \texttt{PxGEQPF}, and in \S \ref{SS=Corrections} we show how to implement the proposed modification. In Section \ref{SS=PZ-bug} we show that for the complex subroutines an additional correction is needed to remove erroneous calls to \texttt{PSCNRM2, PDZNRM2} (for explicit computations of the partial column norms) which also cause bad pivot selections. In this case, the problem is pure programming bug, and in \S \ref{SS=Bug2-discuss} we argue that the probability of detecting it by the usual testing with random matrices is tiny.  Final remarks are given in \S \ref{S=Conclusions}.

%% file: 02_numex.tex
%!TEX root=scalapack_wn.tex

\section{How \texttt{PxGEQPF} can fail and why}\label{S=2}
To illustrate the problem, we run \texttt{PxGEQPF} on a contrived example. We should warn the reader that the example below may not be reproducible on his/her computing platform, and that experimenting with the parameters might be needed to discover instances that exhibit the undesired behavior. 

\subsection{An example}\label{SS=Example}

For the record, in the experiments we have used the following computational environment:
\begin{itemize}
	\item 2x Intel(R) Xeon(R) E5-2690 v3 @ 2.60GHz (24 cores in total);
	\item 256 GB RAM, each processor is equipped with 30 MB of cache memory;
    \item CentOS Linux release 7.6;
	\item Intel Parallel Studio XE 2016 + MKL 11.3;
	\item \texttt{gfortran} 4.8.5 with the built-in system \blas and \lapack libraries;
	\item \scalapack 2.0.2.
\end{itemize}

The following examples are generated by using the \texttt{gfortran} compiler, but the same effects are easily obtained with \texttt{ifort} as well.

\begin{example}
    \label{ex:1}
    As pointed out in \cite{DrmBuj08},  Kahan matrices can be used to quickly find many instances of erroneous pivoting.
    Let $c \in [0, 1]$, $s^2 + c^2 = 1$ with $s>0$, $\mathcal{K}_1(c) = [1]$, and let
    $$
    	\mathcal{K}_n(c) =
    		\left( \begin{array}{c|c}
    			1 & -c \\ \hline
    			0 & s \mathcal{K}_{n-1}(c)
    		\end{array} \right),
    	\quad
    $$
    denote the Kahan matrix of order $n$. Consider the matrix $\mathcal{M}_n(c) = \mathcal{K}_n(c) + \mathcal{K}^\tr_n(c)$, for $n=500$, and $c = 0.44300000000000006$. This matrix was provided as input to the \scalapack routine \texttt{PDGEQPF}, and the diagonal of the upper triangular factor $R$ produced by the routine is shown in Figures \ref{fig:R1a} and \ref{fig:R1b}. 
    
    Figure \ref{fig:R1a} shows the result of running \texttt{PDGEQPF} on a grid divided into \texttt{NPROW=6} process rows and \texttt{NPCOL=4} process columns. Figure \ref{fig:R1b} uses the same input matrix, but now with \texttt{NPROW=4} and \texttt{NPCOL=6}. Both plots demonstrate the failure of the pivoting: the red line showing absolute values of the diagonal elements should be decreasing, and it should stay above the blue line, showing $\max_{j=i+1:n} \|R(i:j, j)\|$, at all times. Furthermore, note that by simply changing the grid topology the result changes, which is unexpected---the input matrix is the same, the compiler and its options are the same, the code being run is the same, although the computation is reordered. With such a behaviour, it is possible that simply upgrading the machine by adding more processors changes the pivoting, the computed factorization, the numerical rank of the same input matrix. This is not a signature of a numerically robust algorithm.
    
%    \todo{Add figure with the run on different topology. Pick the worst case and also (if it happens) correct one.}
    
    \begin{figure}
        \begin{subfigure}{.495\textwidth}
        	\centering
            \includegraphics[width=1\textwidth]{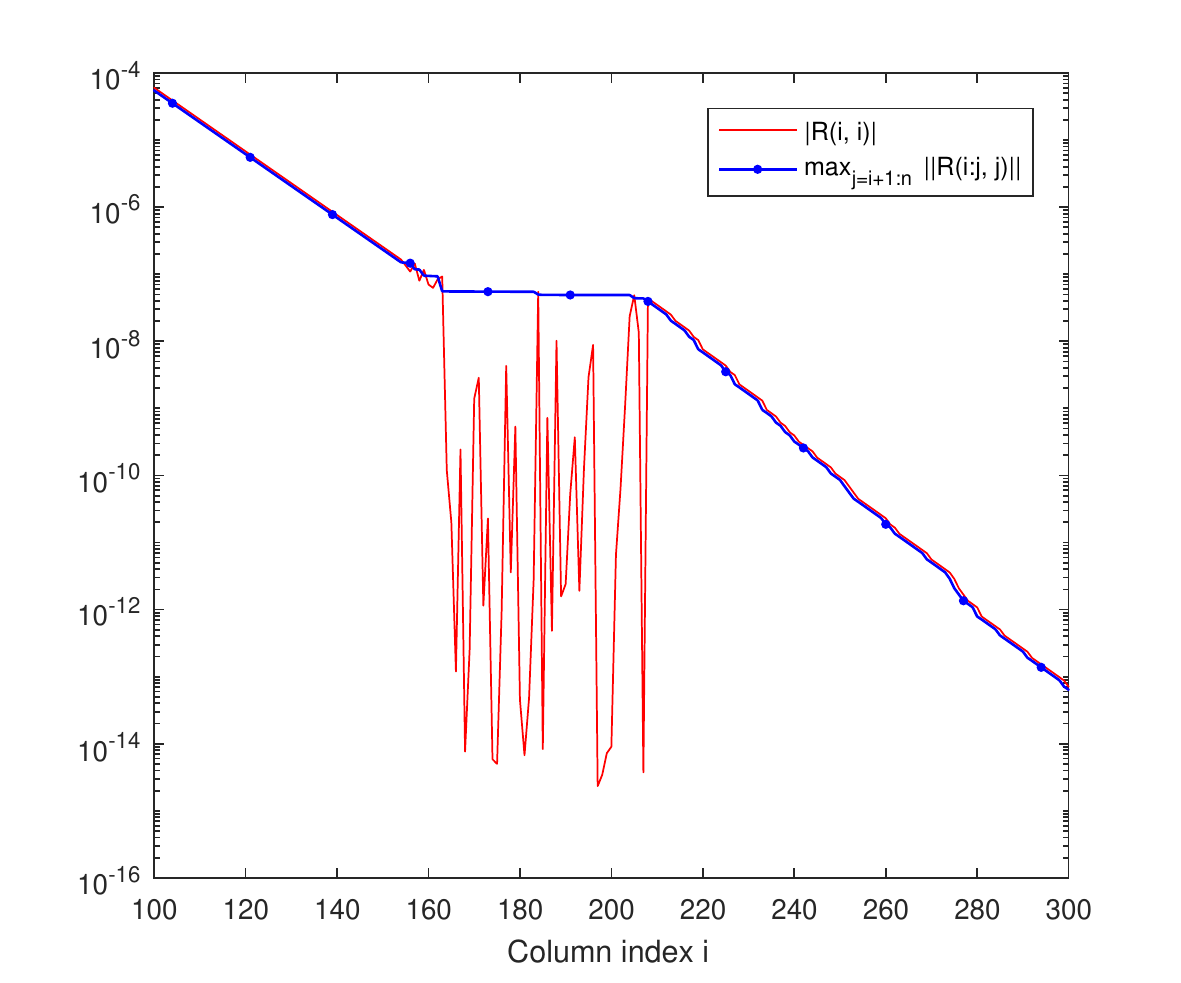}
            \caption{Example \ref{ex:1}, \texttt{NPROW=6} and \texttt{NPCOL=4}}
            \label{fig:R1a}
        \end{subfigure}
        \hfill
        \begin{subfigure}{.495\textwidth}
        	\centering
            \includegraphics[width=1\textwidth]{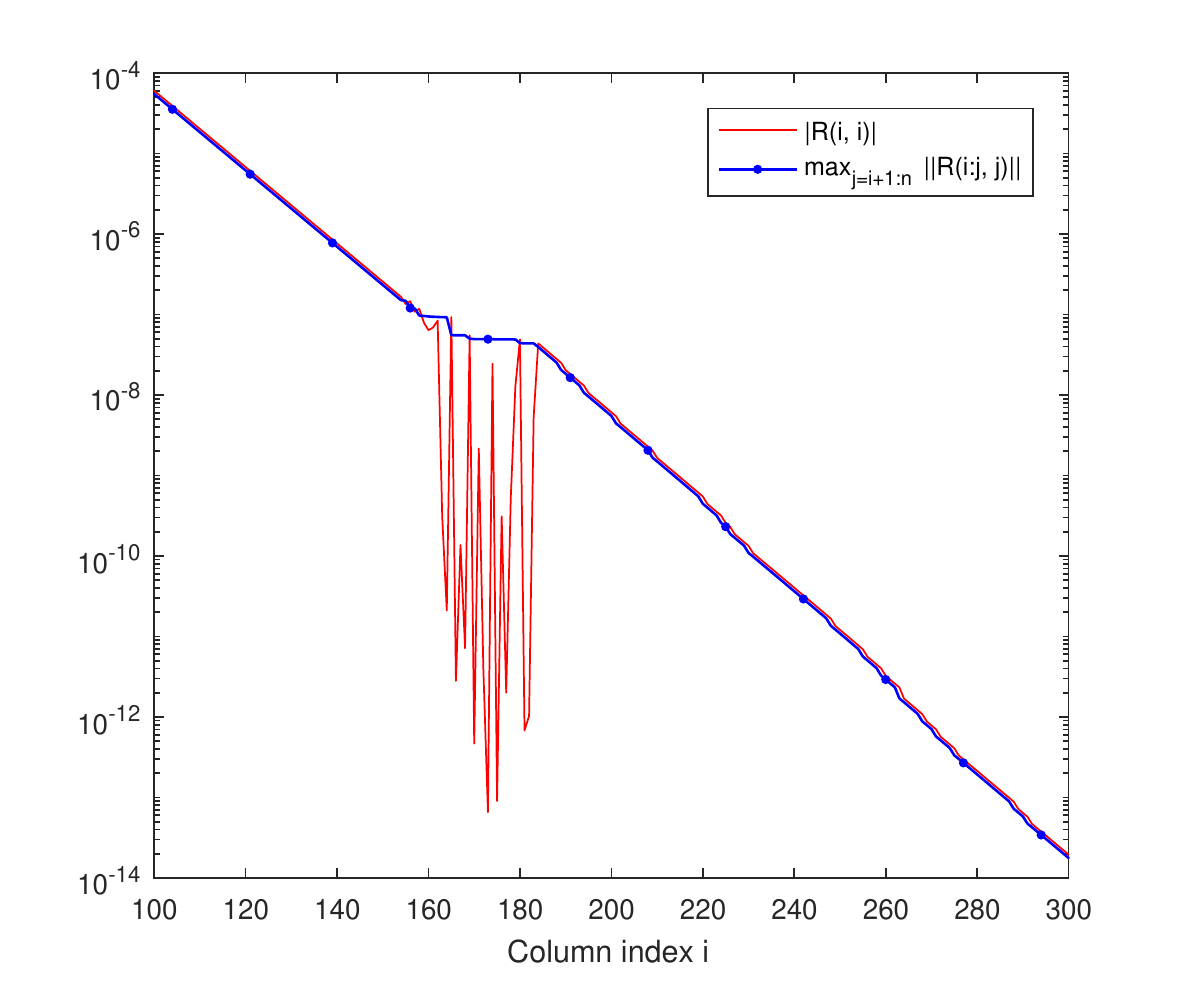}
            \caption{Example \ref{ex:1}, \texttt{NPROW=4} and \texttt{NPCOL=6}}
            \label{fig:R1b}
        \end{subfigure}
        
        \begin{subfigure}{.495\textwidth}
        	\centering
            \includegraphics[width=1\textwidth]{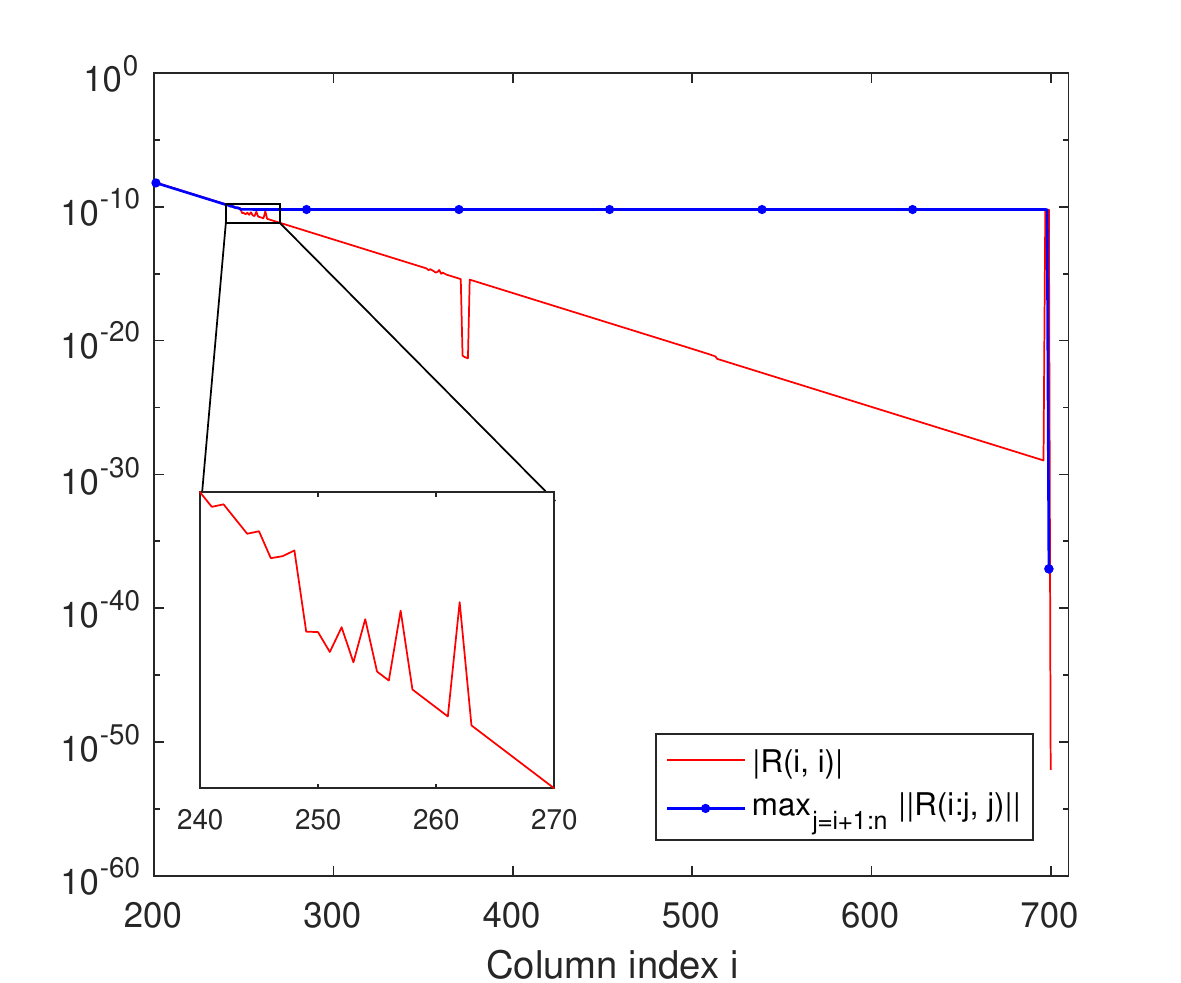}
            \caption{Example \ref{ex:2}, \texttt{NPROW=6} and \texttt{NPCOL=4}}
            \label{fig:R2a}
        \end{subfigure}
        \hfill
        \begin{subfigure}{.495\textwidth}
        	\centering
            \includegraphics[width=1\textwidth]{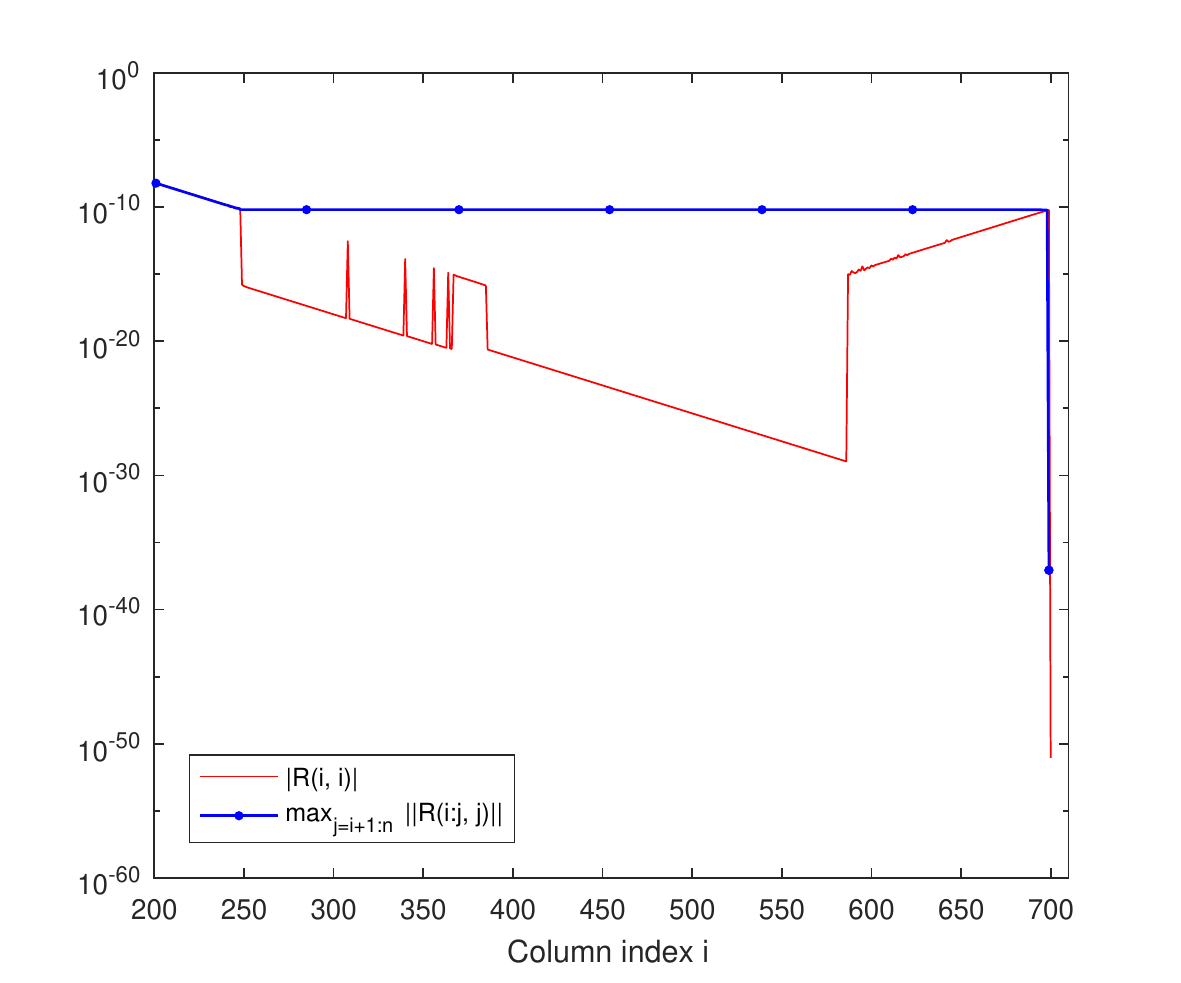}
            \caption{Example \ref{ex:2}, \texttt{NPROW=4} and \texttt{NPCOL=6}}
            \label{fig:R2b}
        \end{subfigure}
        
        \caption{\emph{For the input matrices described in Examples \ref{ex:1} (top row) and \ref{ex:2} (bottom row), absolute values $|R(i, i)|$ of the diagonal elements for the upper-triangular factor $R$ computed by the \texttt{PDGEQPF} routine are shown in red. The blue lines show maximum partial column norms $\max_{j=i+1:n} \|R(i:j, j)\|$. \textbf{A correct algorithm should produce a monotonically descending red line which always stays above the blue line.}
        The plots in the left column show the result when running on a grid configured in \texttt{NPROW=6} process rows and \texttt{NPCOL=4} columns, while the plots in the right column use \texttt{NPROW=4} and \texttt{NPCOL=6}.}}
        \label{fig:R1}
    \end{figure}
\end{example}

The second example shows this effect in an even more drastic way.

\begin{example}
    \label{ex:2}
    Let the input matrix be the Kahan matrix, $\mathcal{K}_n(c)$, for $n=700$ and $c=0.41800000000000004$. In exact arithmetic,  $\mathcal{K}_n(c)$ already has the structure (\ref{QR-structure}) and in (\ref{eq:QRCP}) both $Q$ and $\Pi$ are identities. The results of running the \texttt{PDGEQPF} routine are shown in the bottom row of  Figure \ref{fig:R1}. 

    % \begin{figure}
    %     \begin{subfigure}{.495\textwidth}
    %     	\centering
    %         \includegraphics[width=1\textwidth]{figures/R_fig.2a.pdf}
    %         \label{fig:R2a}
    %     \end{subfigure}
    %     \hfill
    %     \begin{subfigure}{.495\textwidth}
    %     	\centering
    %         \includegraphics[width=1\textwidth]{figures/R_fig.2b.pdf}
    %         \label{fig:R2b}
    %     \end{subfigure}
    %     \caption{For Example \ref{ex:2}, absolute values $|R(i, i)|$ of the diagonal elements for the upper-triangular factor $R$ computed by the \texttt{PDGEQPF} routine are shown in red for $i=200:700$. The blue line shows maximum partial column norms $\max_{j=i+1:n} \|R(i:j, j)\|$, again for $i=200:700$. A correct algorithm should produce a monotonically descending red line which always stays above the blue line. The left plot shows the result when running on a grid configured in \texttt{NPROW=6} process rows and \texttt{NPCOL=4} columns, while the right plot uses \texttt{NPROW=4} and \texttt{NPCOL=6}.}
    %     \label{fig:R2}
    % \end{figure}
\end{example}

Due to the subtlety of the bug, the choice of $c$ producing an erroneous output depends on many factors (compiler, compiler options, processor configuration, etc.), and may differ from the ones used above on a particular system. However, since the presented test matrices are paramatrized by a single parameter, the reader can easily generate similar situations on his/her computer.

This failure to produce the structure (\ref{QR-structure}) makes it very easy to make more mistakes. For instance, since (\ref{QR-structure}) is considered indubitable\footnote{The history of the problem confirms that (\ref{QR-structure}) has always been taken for granted.}, the numerical rank is determined by scanning the diagonal of $R$ downwards from the upper left corner and it is set to $k$ if $k$ is the first index for which $|R_{k+1,k+1}|< \tau |R_{kk}|$, where $\tau$ is a threshold value. This may cause severe underestimating of the numerical rank which then causes, e.g., entirely wrong solution of a least squares problem, or missing many important directions if the factorization is used in computing a POD basis for model order reduction. Further, the important preconditioning effect can be lost. Moreover, changing the processes' topology or adding more processors may change the results considerably!

\subsection{The source of the failure}\label{SS=2.1}
For the sake of completeness and for the reader's convenience, we briefly explain the source of the problem. This will then make the modification of the source code in Section \ref{S=3} clear. For a detailed analysis and discussion we refer the reader to \cite{DrmBuj08}, which is used in this section.

\subsubsection{Partial column norm down-dating}\label{SS=2-pcn=dd}
Consider the $k$--th step in the Householder QR factorization with column pivoting. The input matrix is $A^{(0)}=A=({\bf a}_1,\ldots,{\bf a}_n)\in \mathbb{C}^{m\times n}$
and let $A^{(k)}$ be the intermediate result after $k$ steps. Let $\Pi_k$ be the column permutation matrix that leaves the first $k$ columns unchanged and that in $[A^{(k)}\Pi_k](k+1:m,k+1:n)$ the first column dominates the others in euclidean length -- this is the essence of the Businger-Golub column pivoting. Consider now the matrix
\begin{equation}\label{Eq-QRCP-step}
A^{(k)}\Pi_k \!=\!\left(\begin{array}{cc|cccc}
% \cdot & \cdot & \cdot & \odot & \cdot & \oplus & \cdot  \cr
                     \cdot    & \cdot & \odot & \cdot & \oplus & \cdot  \cr
                               & \cdot & \odot & \cdot & \oplus & \cdot  \cr\hline
                               &       & \circledcirc & \cdot & \circledast & \cdot \cr
                               &       & \circledcirc & \cdot & \ast & \cdot \cr
                               &       & \circledcirc & \cdot & \ast & \cdot \cr
                               &       & \circledcirc & \cdot & \ast & \cdot \cr\end{array}\right),\;\;
%                        &       &       & \circledcirc & \cdot & \ast & \cdot \cr},\;\;
{\bf a}_j^{(k)}\! =\!
\begin{pmatrix}\oplus\cr\oplus\cr\hline\circledast\cr\ast\cr\ast\cr\ast\cr\end{pmatrix}
\!\equiv\! \begin{pmatrix}{\bf x}_j^{(k)} \cr\hline \eta_j^{(k)}\cr {\bf
y}_j^{(k)}\end{pmatrix},\;\; \begin{array}{l} \eta_j^{(k)}\!\!=\circledast \equiv (A^{(k)})_{kj} ,\\
{\bf z}_j^{(k)}\!\!= \begin{pmatrix}\eta_j^{(k)}\cr {\bf
y}_j^{(k)}\end{pmatrix}\in\mathbb{C}^{m-k}\end{array}\;\; j = k+1,\ldots, n.
\end{equation}
To perform the pivoting, the permutation $\Pi_k$ used the norms $\omega_{j}^{(k)}=\|{\bf z}_j^{(k)}\|$, which must be available at each step.
Elements to be annihilated (marked by $\circledcirc$) are  in the vector ${\bf z}_k^{(k)}$.
Let $\mathrm{H}_k$ be
Householder reflector such that
\begin{equation}\label{Hk}
\mathrm{H}_k\begin{pmatrix}\eta_k^{(k)}\cr {\bf
y}_k^{(k)}\end{pmatrix}=\begin{pmatrix}R_{kk}\cr 0\end{pmatrix}, \;\mbox{and let, for $j>k$},\;\;
\begin{pmatrix}\beta_j^{(k+1)}\cr {\bf z}_j^{(k+1)}\end{pmatrix}
%\equiv
%\pmatrix{\beta_j^{(k+1)}\cr \eta_j^{(k+1)}\cr {\bf y}_j^{(k+1)}}
=\mathrm{H}_k {\bf z}_j^{(k)} .
%\pmatrix{\eta_j^{(k)}\cr {\bf y}_j^{(k)}}.
\end{equation}
Set $Q_k = I_k \oplus H_k$ and compute the next iteration
\begin{equation}\label{Eq-QRCP-step-1}
A^{(k+1)} = Q_k^* A^{(k)}\Pi_k \!=\!\left(\begin{array}{ccc|ccc}
% \cdot & \cdot & \cdot & \odot & \cdot & \oplus & \cdot  \cr
                     \cdot    & \cdot & \odot & \cdot & \oplus & \cdot  \cr
                               & \cdot & \odot & \cdot & \oplus & \cdot  \cr
                               &       & \blacksquare & \cdot & \circledast & \cdot \cr\hline
                               &       & 0 & \cdot & \ast & \cdot \cr
                               &       & 0 & \cdot & \ast & \cdot \cr
                               &       & 0 & \cdot & \ast & \cdot \cr\end{array}\right),\;\;
%                        &       &       & \circledcirc & \cdot & \ast & \cdot \cr},\;\;
{\bf a}_j^{(k+1)}\! =\!
\begin{pmatrix}\oplus\cr\oplus\cr\circledast\cr\hline\ast\cr\ast\cr\ast\cr\end{pmatrix}
\!\equiv\! \begin{pmatrix}{\bf x}_j^{(k+1)} \cr \hline \eta_j^{(k+1)}\cr {\bf
y}_j^{(k+1)}\end{pmatrix},\;\; \begin{array}{l} \eta_j^{(k+1)}\!\!=\circledast \equiv (A^{(k+1)})_{k+1,j} ,\\
{\bf z}_j^{(k+1)}\!\!= \begin{pmatrix}\eta_j^{(k+1)}\cr {\bf
y}_j^{(k+1)}\end{pmatrix}\in\mathbb{C}^{m-k-1}\\ j=k+2,\ldots ,n.\end{array}  
\end{equation}
The permutation $\Pi_k$ ensures that $\blacksquare = |R_{kk}|\geq
\omega_{j}^{(k)}$ for all $j\geq k$. 
%%%(In the blocked algorithm, $\mathrm{H_k}$ is replaced with aggregated reflectors.)
For the next step, in order to determine the permutation $\Pi_{k+1}$, we need the column norms $\omega_j^{(k+1)}= \|{\bf z}_j^{(k+1)}\|$. 
To that end, we recall (\ref{Eq-QRCP-step}) and (\ref{Hk}).
Orthogonality of $\mathrm{H}_k$ implies that in (\ref{Hk}) the norm of ${\bf z}_j^{(k)}$ equals
$\omega_j^{(k)}=\sqrt{(\beta_j^{(k+1)})^2+ \|{\bf
z}_j^{(k+1)}\|^2}$, and thus
\begin{equation}\label{LAPACK-update}
\omega_j^{(k+1)}=\sqrt{(\omega_j^{(k)})^2 - (\beta_j^{(k+1)})^2 }=
\omega_j^{(k)}\sqrt{1-\left(\frac{\beta_j^{(k+1)}}{\omega_j^{(k)}}\right)^2}.
\end{equation}
Since initially $\omega_j^{(1)}=\|{\bf a}_j\|$, each $\omega_j^{(k+1)}$ can be recursively computed from
$\omega_j^{(k)}$ and $\beta_j^{(k+1)}$, using (\ref{LAPACK-update}). This is appealing because recomputing the column norms of the trailing submatrices of the $A^{(k)}$'s incurs an unacceptable increase of computational complexity. It also facilitates use of the aggregated transformations, because only the $\beta_j^{k+1}$'s are needed to compute the next pivoting columns; this is used in \texttt{xGEQP3}.

\subsubsection{Massive cancellations and safety switch}\label{SS=MCSS}
Note that $(\omega_j^{(k)})_{k\geq 1}$ is nonincreasing sequence,
obtained by successive subtractions, which makes it prone to multiple catastrophic cancellations. In \linpack (and later in \lapack and many other software packages) a safety device monitors the down-dating history of each partial column norm and, 
if at some step $k$ the update (\ref{LAPACK-update}) is not
considered to be numerically safe, the corresponding value
$\widetilde\omega_j^{(k+1)}$ is computed explicitly by calling a function that computes the euclidean vector norm. 
Note that we use the tilde \ $\widetilde{}$ \ to denote actually computed quantities. 
After this explicit norm computation, a copy of 
$\widetilde\omega_j^{(k+1)}$ is  stored in the
variable $\widetilde\nu_j$, $\widetilde\nu_j=\widetilde\omega_j^{(k+1)}$. Thus, at
any moment in the algorithm, $\widetilde\nu_j$ contains the last
explicitly computed partial column norm in the $j$--th column.  Initially, 
$\widetilde\nu_j=computed(\|A(:,j)\|)$.

The safety device in \lapack and \scalapack first computes the control variables 
%which allows using update by (\ref{LAPACK-update})
%has simple and elegant structure:
\begin{equation}\label{lapack-update-structure}
 \texttt{TEMP} = computed(\underbrace{\left( 1 - \left(\frac{\widetilde\beta_j^{(k+1)}}
 {\widetilde\omega_j^{(k)}}\right)^2\right)}_{predicted\;\; loss}\;\; ; \;\;\;\;
 % > \tol,\;\; \tol \approx 20 \roff, \\
 \texttt{TEMP2} = computed(1 + 0.05\cdot  \texttt{TEMP} \cdot\!\!\!\! \underbrace{\left(\frac{\widetilde\omega_j^{(k)}}{\widetilde\nu_j}\right)^2
 }_{memorized\;\; loss}\!\!\!\! )
 \end{equation}
 where the {\em predicted loss} part ($\approx (\widetilde\omega_j^{(k+1)}/\widetilde\omega_j^{(k)})^2$) estimates loss of accuracy in computing $\widetilde\omega_j^{(k+1)}$
from $\widetilde\omega_j^{(k)}$, and the {\em memorized loss} part memorizes the cumulative
loss of accuracy (by cancellations) since the last update by explicit norm computation.
The two factors multiplied together indicate how accurately $\widetilde\omega_j^{(k+1)}$ approximates
the corresponding partial column norm, i.e. how much  the norm has dropped by the subtractions (\ref{LAPACK-update}) since its last explicit computation. Then, $\texttt{TEMP2}$ is compared to one: if it equals one, then the norm is recomputed explicitly; otherwise the formula (\ref{LAPACK-update}) is deployed.\\

\begin{algorithm}
\caption{The partial column norm down-dating strategy in \linpack/\lapack/\scalapack \label{ALG:1}}
\begin{algorithmic}[1]
% \STATE $\texttt{TEMP} = computed(\underbrace{\left( 1 - \left(\frac{\widetilde\beta_j^{(k+1)}}
% {\widetilde\omega_j^{(k)}}\right)^2\right)}_{predicted}$ ; 
% $\texttt{TEMP2} = computed(1 + 0.05\cdot  \texttt{TEMP} \cdot \underbrace{\left(\frac{\widetilde\omega_j^{(k)}}{\widetilde\nu_j}\right)^2
% }_{memorized})$
\STATE Compute \texttt{TEMP2} as in (\ref{lapack-update-structure}).
\IF{\texttt{TEMP2} == 1}
\STATE Compute $\widetilde\omega_j^{(k+1)}$ by explicit vector norm computation and also set $\widetilde{\nu_j}=\widetilde\omega_j^{(k+1)}$.
\ELSE
\STATE Compute  $\widetilde\omega_j^{(k+1)}$using the formula (\ref{LAPACK-update}).
\ENDIF
\end{algorithmic}
\end{algorithm}

The testing of $\texttt{TEMP2}$ against one was probably meant by the developers of \texttt{xqrdc} to test 
\begin{equation}\label{eq:eps-test}
\mbox{whether}\:\;0.05\cdot  \texttt{TEMP} \cdot \left(\frac{\widetilde\omega_j^{(k)}}{\widetilde\nu_j}\right)^2 < \roff, \;\; \mbox{i.e. whether (roughly)} \;\; \left(\frac{\widetilde\omega_j^{(k+1)}}{\widetilde\nu_j}\right)^2 < 20 \roff ,
\end{equation}
where $\roff$ denotes the roundoff unit.

\subsubsection{Discussion}\label{SS=modif}
Using the comparison in Line 2 of Algorithm \ref{ALG:1} to check (\ref{eq:eps-test}) 
is problematic if at that moment the variable $\texttt{TEMP2}$ is computed in a long register on the CPU -- the extra precision precludes detecting the (intended) critical level of the (implicitly) tested value.  Long registers provide extra precision which is invaluable in finite precision computations. However, it should be used with great care and compiler manuals point to this delicate issue, as e.g. in the following two examples.

\begin{verbatim}
https://gcc.gnu.org/onlinedocs/gcc-3.1.1/gcc/index.html#Top

-ffloat-store
    Do not store floating point variables in registers, and inhibit other options that 
    might change whether a floating point value is taken from a register or memory.

    This option prevents undesirable excess precision on machines such as the 68000 where 
    the floating registers (of the 68881) keep more precision than a double is supposed 
    to have. Similarly for the x86 architecture. For most programs, the excess precision 
    does only good, but a few programs rely on the precise definition of IEEE floating 
    point. Use -ffloat-store for such programs, after modifying them to store all 
    pertinent intermediate computations into variables. 
\end{verbatim}
    
\begin{verbatim}
https://www.nag.co.uk/nagware/np/r62_doc/manual/compiler_2_4.html   
NAG Fortran Compiler, Release 6.2

-float-store
    (Gnu C based systems only) Do not store floating-point variables in registers on 
    machines with floating-point registers wider than 64 bits. This can avoid problems 
    with excess precision. 
\end{verbatim}
In \cite{DrmBuj08} we show that this is indeed an important issue in the pivoted QR factorization codes.  Simply by invoking this compiler option, thus preventing the comparison of the long register value of \texttt{TEMP2} with one,  may considerably change the computed factorization. 

The same undesirable effect is obtained if immediately after computing \texttt{TEMP2}, and before comparing it with one, we insert a write statement to display the value of \texttt{TEMP2} (\texttt{WRITE(*,*) TEMP2}). The  \texttt{WRITE} command causes spilling \texttt{TEMP2} to working precision memory location, thus possibly changing its value, and the result of  Line 2 of Algorithm \ref{ALG:1} might be different. We refer the reader to \cite{DrmBuj08}, \cite{SLWN-2010-1}, where numerous examples are given how a \texttt{WRITE(*,*)} statement dramatically changes the computed numerical rank, the solution of a least squares problem, or the staircase form of a linear time invariant dynamical system.

The above problem can be removed by replacing the implicit test in Line 2 with the scheme outlined in Algorithm \ref{ALG:2} .
\begin{algorithm}
\caption{A modified column norm down-dating strategy \label{ALG:2}}
\begin{algorithmic}[1]
% \STATE $\texttt{TEMP} = computed(\underbrace{\left( 1 - \left(\frac{\widetilde\beta_j^{(k+1)}}
% {\widetilde\omega_j^{(k)}}\right)^2\right)}_{predicted}$ ; 
% $\texttt{TEMP2} = computed(1 + 0.05\cdot  \texttt{TEMP} \cdot \underbrace{\left(\frac{\widetilde\omega_j^{(k)}}{\widetilde\nu_j}\right)^2
% }_{memorized})$
\STATE $\texttt{TEMP2}  = \texttt{TEMP} \cdot \left({\widetilde\omega_j^{(k)}}/{\widetilde\nu_j}\right)^2$. \COMMENT{Note that this is different from (\ref{lapack-update-structure})}
\IF{\texttt{TEMP2} $\leq tol$}
\STATE Compute $\widetilde\omega_j^{(k+1)}$ by explicit vector norm computation and also set $\widetilde{\nu_j}=\widetilde\omega_j^{(k+1)}$.
\ELSE
\STATE Compute  $\widetilde\omega_j^{(k+1)}$using the formula (\ref{LAPACK-update}).
\ENDIF
\end{algorithmic}
\end{algorithm}
If we want this to be compatible with (\ref{eq:eps-test}), then $tol = 20\roff$. This modified switching between the scalar formula down-dating and explicit norm computation improves the result on many examples, but not all of them. There is still a possibility that to the pivoting device a column may appear of much larger norm that it actually is and may be wrongly selected as a new pivot. If that occurs for several columns then the result of pivoting can be such as shown on the figures in \S \ref{SS=Example}. Simply put, the pivoting procedure is in the dark completely about the actual norms of the columns among which the pivots are selected. (Actually, even a zero column could be selected as pivot despite the fact that all remaining pivot candidates are nonzero.)

A tedious analysis in \cite{DrmBuj08} shows that the proper tolerance level in Line 2 of Algorithm \ref{ALG:2} is $tol=\sqrt{\roff}$. 
In the next section, we show how to implement this modification in the source code of \texttt{PxGEQPF}.

%% file: 03_correction.tex
%!TEX root=scalapack_wn.tex

\section{New version of \texttt{PxGEQPF}}\label{S=3}
In this section we show to implement the modification from \S \ref{SS=modif} in a backward compatible way. The changes will be explained using the critical parts of the source code of \texttt{PDGEQPF}. 
% The source of the error of the \pdgeqpf routine lies in the update of the column norms, i.e., in the criterion of when this norm should be recomputed from scratch instead of being updated in a more efficient, but potentially numerically unstable way. Below is the critical part of the routine:

First, define the tolerance  \tolthreez  as a double precision variable
\begin{minted}[linenos, bgcolor=mintedbg, frame=lines]{fortranfixed}
TOL3Z = SQRT( DLAMCH('Epsilon') )
\end{minted}

\subsection{Critical parts of the source code of \texttt{PDGEQPF}}
\noindent The critical parts of the code are \\

\framebox{\textbf{\texttt{PDGEQPF.F}, lines 479:497}}
\begin{minted}[linenos, firstnumber=479, bgcolor=mintedbg, frame=lines]{fortranfixed}
               IF( WORK( IPN+LL ).NE.ZERO ) THEN
                  TEMP = ONE-( ABS( WORK( IPW+LL ) ) /
     $                         WORK( IPN+LL ) )**2
                  TEMP = MAX( TEMP, ZERO )
                  TEMP2 = ONE + 0.05D+0*TEMP*
     $                    ( WORK( IPN+LL ) / WORK( IPN+NQ+LL ) )**2
                  IF( TEMP2.EQ.ONE ) THEN
                     IF( IA+M-1.GT.I ) THEN
                        CALL PDNRM2( IA+M-I-1, WORK( IPN+LL ), A, I+1,
     $                               J+LL-JJ+2, DESCA, 1 )
                        WORK( IPN+NQ+LL ) = WORK( IPN+LL )
                     ELSE
                        WORK( IPN+LL ) = ZERO
                        WORK( IPN+NQ+LL ) = ZERO
                     END IF
                  ELSE
                     WORK( IPN+LL ) = WORK( IPN+LL ) * SQRT( TEMP )
                  END IF
               END IF
\end{minted}

% A similar column norm update appears in lines $509$--$525$ of the same file:
\noindent and \\

\framebox{\textbf{\texttt{PDGEQPF.F}, lines 508:526}}

\begin{minted}[linenos, firstnumber=508, bgcolor=mintedbg, frame=lines]{fortranfixed}
                  IF( WORK( IPN+LL ).NE.ZERO ) THEN
                     TEMP = ONE-( ABS( WORK( IPW+LL ) ) /
     $                            WORK( IPN+LL ) )**2
                     TEMP = MAX( TEMP, ZERO )
                     TEMP2 = ONE + 0.05D+0*TEMP*
     $                     ( WORK( IPN+LL ) / WORK( IPN+NQ+LL ) )**2
                     IF( TEMP2.EQ.ONE ) THEN
                        IF( IA+M-1.GT.I ) THEN
                           CALL PDNRM2( IA+M-I-1, WORK( IPN+LL ), A,
     $                                  I+1, K+LL-JJ+1, DESCA, 1 )
                           WORK( IPN+NQ+LL ) = WORK( IPN+LL )
                        ELSE
                           WORK( IPN+LL ) = ZERO
                           WORK( IPN+NQ+LL ) = ZERO
                        END IF
                     ELSE
                        WORK( IPN+LL ) = WORK( IPN+LL ) * SQRT( TEMP )
                     END IF
                  END IF
\end{minted}

\subsection{Proposed corrections}\label{SS=Corrections}

\noindent The proposed correction replaces the above lines with, respectively, \\

\framebox{\textbf{\texttt{PDGEQPF.F}, modified routine, lines 479:495}}

\begin{minted}[linenos, firstnumber=479, bgcolor=mintedbg, frame=lines]{fortranfixed}
               IF( WORK( IPN+LL ).NE.ZERO ) THEN
                  TEMP = ABS( WORK( IPW+LL ) ) / WORK( IPN+LL )
                  TEMP = MAX( ZERO, ( ONE+TEMP )*( ONE-TEMP ) )
                  TEMP2 = TEMP*( WORK( IPN+LL ) / WORK( IPN+NQ+LL ) )**2
                  IF( TEMP2.LE.TOL3Z ) THEN
                     IF( IA+M-1.GT.I ) THEN
                        CALL PDNRM2( IA+M-I-1, WORK( IPN+LL ), A, I+1,
     $                               J+LL-JJ+2, DESCA, 1 )
                        WORK( IPN+NQ+LL ) = WORK( IPN+LL )
                     ELSE
                        WORK( IPN+LL ) = ZERO
                        WORK( IPN+NQ+LL ) = ZERO
                     END IF
                  ELSE
                     WORK( IPN+LL ) = WORK( IPN+LL ) * SQRT( TEMP )
                  END IF
               END IF
\end{minted}

\noindent and\\

\framebox{\textbf{\texttt{PDGEQPF.F}, modified routine, lines 508:525}}

\begin{minted}[linenos, firstnumber=508, bgcolor=mintedbg, frame=lines]{fortranfixed}
                  IF( WORK( IPN+LL ).NE.ZERO ) THEN
                     TEMP = ABS( WORK( IPW+LL ) ) / WORK( IPN+LL )
                     TEMP = MAX( ZERO, ( ONE+TEMP )*( ONE-TEMP ) )
                     TEMP2 = TEMP*
     $                         ( WORK( IPN+LL ) / WORK( IPN+NQ+LL ) )**2
                     IF( TEMP2.LE.TOL3Z ) THEN
                        IF( IA+M-1.GT.I ) THEN
                           CALL PDNRM2( IA+M-I-1, WORK( IPN+LL ), A,
     $                                  I+1, K+LL-JJ+1, DESCA, 1 )
                           WORK( IPN+NQ+LL ) = WORK( IPN+LL )
                        ELSE
                           WORK( IPN+LL ) = ZERO
                           WORK( IPN+NQ+LL ) = ZERO
                        END IF
                     ELSE
                        WORK( IPN+LL ) = WORK( IPN+LL ) * SQRT( TEMP )
                     END IF
                  END IF
\end{minted}

\begin{remark}
This modification was theoretically analyzed and tested in \cite{DrmBuj08}.
In \cite[\S 4.3]{DrmBuj08} we also proposed a stronger (in the sense of error analysis) partial column norm down-dating scheme that needed an extra $n$-dimensional array in the work space. This extra work space precluded backward compatibility, and was not used in the modifications of \texttt{xGEQPF} and \texttt{xGEQP3}.
\end{remark}

\begin{remark}
Similar modification can be applied to 
\texttt{xGEQPX} in  \cite{bischof-q-orti-RRQR-1998-TOMS782}, and make the corresponding rank revealing strategy more robust. 
\end{remark}

\begin{remark}\label{RE-discovery}
We discovered the problem through rigorous stress testing of the Jacobi SVD algorithm \cite{drm-ves-VW-1, drm-ves-VW-2} which uses (\ref{eq:QRCP}, \ref{QR-structure}) in the pre-processing phase as a preconditioner for the one sided Jacobi iterations. Systematic large scale adversarial testing was used to check the theoretical error bounds, in particular when the input matrices only barely satisfied  the assumptions of the perturbation theory. The testing procedure singled out all matrices for which the measured error in the singular values was larger than predicted by the perturbation theory. After checking the stored control variables, we discovered that in all those suspicious cases the row scaled matrix $R_r = \mathrm{diag}(1/\|R(i,:)\|)_{i=1}^n R$ was extremely ill-conditioned, which we knew it shouldn't be happening, because  of the diagonal dominance (\ref{QR-structure}).
\end{remark}

\section{Another error in  \texttt{PCGEQPF} and \texttt{PZGEQPF}}\label{SS=PZ-bug}
After successful testing of the modification described in \S  \ref{SS=Corrections} on the real data (\texttt{PSGEQPF}, \texttt{PDGEQPF}) we routinely changed the complex subroutines \texttt{PCGEQPF} and \texttt{PZGEQPF}. Unfortunately and unexpectedly, the complex subroutines did not pass the test. The failure was of the same kind -- the pivoting was wrong, despite our modification!

\subsection{An example of failure}
The complex versions of \texttt{PxGEQPF} seem to contain one additional error in the norm-updating part of the code, unrelated to the numerical issue discussed above. In fact, we were able to trace the error to the branch of the down-dating strategy where the partial column norm is computed explicitly by calling the \texttt{PyxNRM2} function.

Consider the original code in \texttt{PCGEQPF}: \\

\framebox{\textbf{\texttt{PCGEQPF.F}, lines 502:520}}

\begin{minted}[linenos, firstnumber=502, bgcolor=mintedbg, frame=lines]{fortranfixed}
            DO 90 LL = JJ, JJ + JN - J - 1
               IF( RWORK( LL ).NE.ZERO ) THEN
                  TEMP = ONE-( ABS( WORK( LL ) ) / RWORK( LL ) )**2
                  TEMP = MAX( TEMP, ZERO )
                  TEMP2 = ONE + 0.05E+0*TEMP*
     $                    ( RWORK( LL ) / RWORK( NQ+LL ) )**2
                  IF( TEMP2.EQ.ONE ) THEN
                     IF( IA+M-1.GT.I ) THEN
                        CALL PSCNRM2( IA+M-I-1, RWORK( LL ), A,
     $                                I+1, J+LL-JJ, DESCA, 1 )
                        RWORK( NQ+LL ) = RWORK( LL )
                     ELSE
                        RWORK( LL ) = ZERO
                        RWORK( NQ+LL ) = ZERO
                     END IF
                  ELSE
                     RWORK( LL ) = RWORK( LL ) * SQRT( TEMP )
                  END IF
               END IF
\end{minted}

\begin{figure}
    \begin{subfigure}{.495\textwidth}
    	\centering
        \includegraphics[width=1\textwidth]{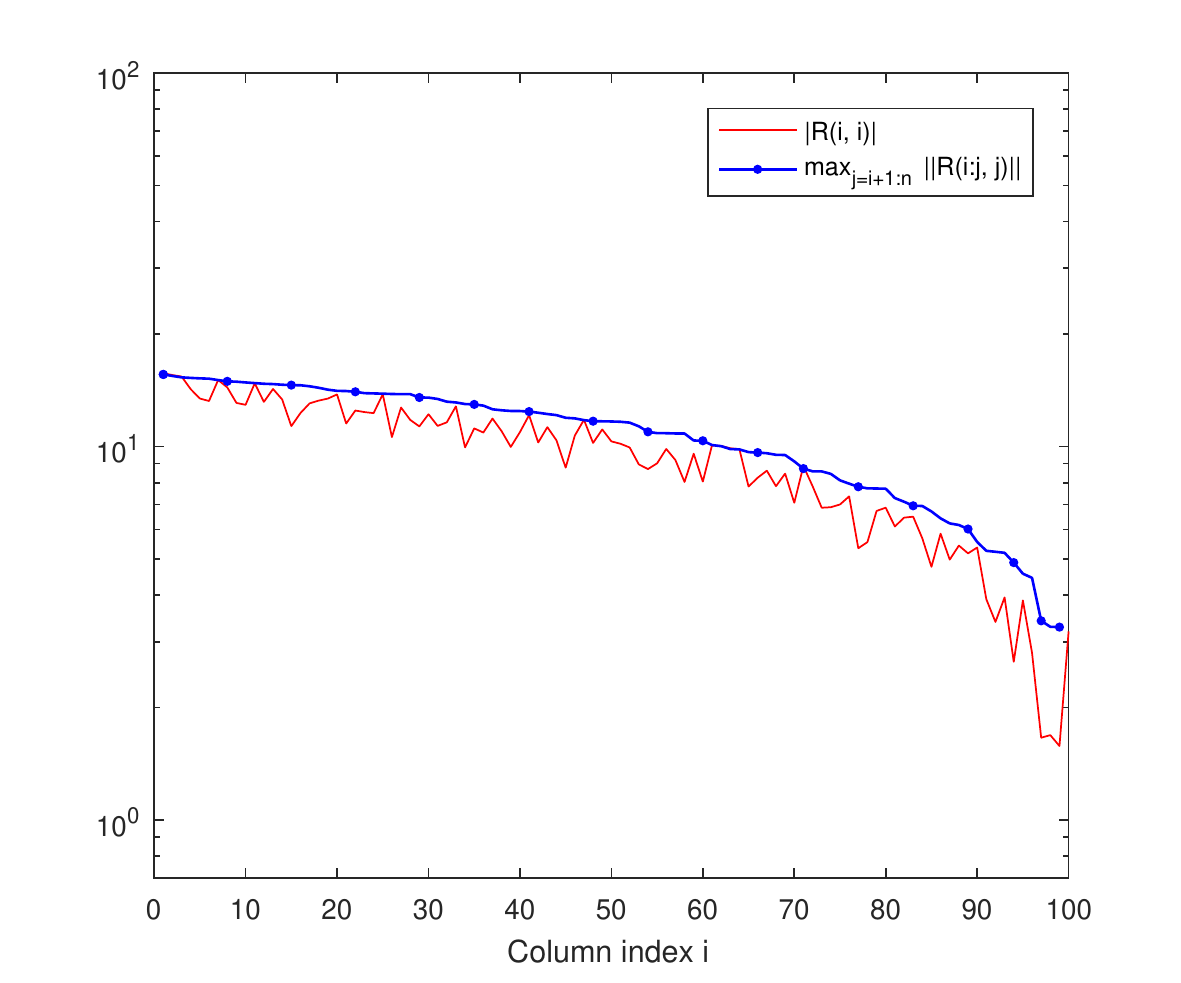}
        \label{fig:R3a}
    \end{subfigure}
    \hfill
    \begin{subfigure}{.495\textwidth}
    	\centering
        \includegraphics[width=1\textwidth]{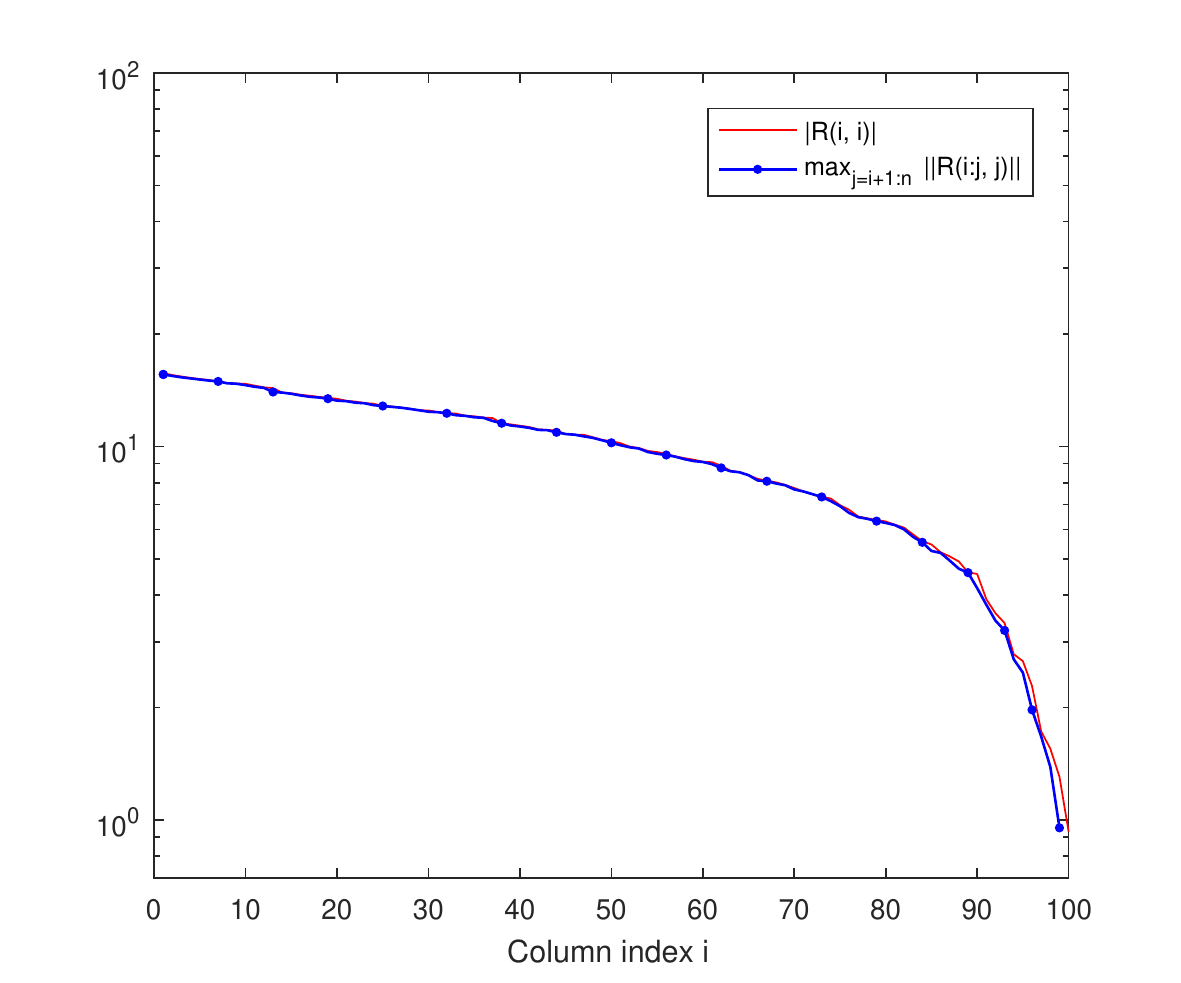}
        \label{fig:R3b}
    \end{subfigure}
    \caption{\emph{The column indexing bug in \texttt{PCGEQPF}. The plot on the left  shows absolute values of the diagonal elements of $R$ (red line) and the partial column norms (blue line) when the original \scalapack{} routine is forced to recompute the partial column norms in each step. \textbf{Recall that the red line should be monotonically decreasing and that it should be above the blue line.} The plot on the right shows the correct output obtained by changing the column index \texttt{J+LL-JJ} to \texttt{J+LL-JJ+1} in Line 510 of the routine. A random $100 \times 100$ matrix was used as input.}}
    \label{fig:R3}
\end{figure}

\noindent To show that the problem is related to the call of \texttt{PSCNRM2} in Line 510, we replace 
Line 508  with

\begin{minted}[linenos, , firstnumber=508, bgcolor=mintedbg, frame=lines]{fortranfixed}
                  IF( ONE.EQ.ONE ) THEN
\end{minted}
 
\noindent thus enforcing explicit call to \texttt{PSCNRM2} at every step. Since in that case there is no down-dating issue,  we expect that the QR-routine never fails to produce $R$ satisfying \eqref{QR-structure}. However, running the code on a random $100 \times 100$ matrix---or, essentially on any matrix---produces a non-sorted diagonal. Figure \ref{fig:R3} demonstrates the issue.

After an analysis, it appears that \texttt{PSCNRM2} is computing the norm of a wrong column: instead of \texttt{J+LL-JJ}, the column index in Line 511 should be \texttt{J+LL-JJ+1}:

\begin{minted}[linenos, firstnumber=510, bgcolor=mintedbg, frame=lines]{fortranfixed}
                        CALL PSCNRM2( IA+M-I-1, RWORK( LL ), A,
     $                                I+1, J+LL-JJ+1, DESCA, 1 )
\end{minted}

\subsection{Corrected code}
Hence, the modified critical part of \texttt{PCGEQPF} reads:\\

\framebox{\textbf{\texttt{PCGEQPF.F}, modified routine, lines 502:519}}

\begin{minted}[linenos, firstnumber=502, bgcolor=mintedbg, frame=lines]{fortranfixed}
            DO 90 LL = JJ, JJ + JN - J - 1
               IF( RWORK( LL ).NE.ZERO ) THEN
                  TEMP = ABS( WORK( LL ) ) / RWORK( LL )
                  TEMP = MAX( ZERO, ( ONE+TEMP )*( ONE-TEMP ) )
                  TEMP2 = TEMP * ( RWORK( LL ) / RWORK( NQ+LL ) )**2
                  IF( TEMP2.LE.TOL3Z ) THEN
                     IF( IA+M-1.GT.I ) THEN
                        CALL PSCNRM2( IA+M-I-1, RWORK( LL ), A,
     $                                I+1, J+LL-JJ+1, DESCA, 1 )
                        RWORK( NQ+LL ) = RWORK( LL )
                     ELSE
                        RWORK( LL ) = ZERO
                        RWORK( NQ+LL ) = ZERO
                     END IF
                  ELSE
                     RWORK( LL ) = RWORK( LL ) * SQRT( TEMP )
                  END IF
               END IF
\end{minted}

After this modification, the complex subroutines have passed all tests.

\subsection{Source of the error}
The programming error is related to a shift of the indices in the complex routines. Presumably, the code for \texttt{PCGEPQF} was created by adapting the routine \texttt{PSGEQPF} which operates on input matrices of type \texttt{REAL}. Unlike the later routine, which uses only a single real work array, the complex routine splits the auxiliary work arrays in two parts: array \texttt{RWORK} of type \texttt{REAL}, and array \texttt{WORK} of type \texttt{COMPLEX}. To simplify indexing in these two arrays, the loop in Line\footnote{Here all line numbers refer to the original subroutine, before modifications.} 502 starts with \texttt{LL=JJ}, while the same loop in \texttt{PSGEQPF} starts with \texttt{LL=JJ-1}. It appears that all the array indices have been correctly updated to reflect this shifting, except for the column index in Line 511: from \texttt{J+LL-JJ+2} in \texttt{PSGEQPF}, it was erroneously translated to \texttt{J+LL-JJ}, while the correct value should be \texttt{J+LL-JJ+1}.

\subsection{Discussion: why has this not been detected by the test routines}\label{SS=Bug2-discuss}

Why is this problem difficult to detect by testing the code on many random matrices of different sizes,  with varying input parameters, process topology, block sizes, compiler options? 

There are three key factors that conspire to make the problem practically undetectable by large scale testing on random matrices. 

First, random matrices are well conditioned with high probability. 
For instance \cite[Theorem 4.6, Theorem 5.6]{Cond-Gauss-Rand-2005} states that for $n\geq m\geq 2$, and a random matrix $A\in\mathbb{C}^{m\times n}$ whose elements are independent and identically distributed standard complex normal random variables, the condition number of $A$ can be estimated in probability by  
\begin{equation}\label{ZJJ1}
 \frac{1}{2\pi}\left( \frac{c}{x}\right)^{2(n-m+1)}  <  \mathbf{P}\left( \frac{\kappa_2(A)}{n/(n-m+1)} > x  \right) \leq \frac{1}{2\pi}\left( \frac{C}{x}\right)^{2(n-m+1)} ,
\end{equation}
where $x\geq n-m+1$ and $C\leq 6.298$, $c\geq 0.319$ are universal constants (independent of $x, m, n$). Further, we have in expectation
\begin{equation}\label{ZJJ2}
    \mathbb{E}[\log\kappa_2(A)] < \log\frac{n}{n-m+1} + 2.240,
\end{equation}
see
\cite[Theorem 6.2]{Cond-Gauss-Rand-2005} and \cite{Edelman:1988:ECN:58846.58854}. Hence, a typically used random matrix to test the code is always expected to be well conditioned.

Secondly, it has been shown in \cite{DrmBuj08} that for a tall\footnote{If $A$ is tall, $m>n$, then apply (\ref{ZJJ1}, \ref{ZJJ2}) to $A^*$ and use $\kappa_2(A)=\kappa_2(A^*)$. Also, the analysis from \cite{DrmBuj08} can be easily adapted for wide matrices, $n>m$, but here we omit those technical details.} matrix $A$ a necessary condition for the failure of the original \linpack down-dating formula is that $\|A_c^\dagger\|\equiv 1/\sigma_{\min}(A_c) > 1/\sqrt{\roff}$, where $A = A_c D$, $D=\mathrm{diag}(\|A(:,i)\|)_{i=1}^n$. In that case we also have that
$\kappa_2(A_c)\geq \|A_c^\dagger\|> 1/\sqrt{\roff}$.

And thirdly, by \cite{slu-69}, 
\begin{equation}
    \kappa_2(A_c) \leq \sqrt{n}\min_{\Delta=\mathrm{diag}}\kappa_2(A\Delta) \leq
    \sqrt{n}\kappa_2(A).
\end{equation}
In general, it possible that $\kappa_2(A_c)\ll\kappa_2(A)$, but in this case of randomly generated $A$, it is most likely that $\kappa_2(A_c)\approx \kappa_2(A)$.
Altogether, as a corollary of the above, we have 
\begin{equation}
    \mathbf{P}\left( \frac{\kappa_2(A_c)/\sqrt{n}}{m/(|n-m|+1)} \leq x  \right) \geq 1 - \frac{1}{2\pi}\left( \frac{C}{x}\right)^{2(|n-m|+1)}.
\end{equation}
Hence, the sensitive branch of the down-dating strategy from \S \ref{SS=2-pcn=dd} and \S \ref{SS=MCSS}, that caused the failure of the pivoting in the ill-conditioned cases, was working, with high probability, well on random matrices (without massive catastrophic cancellations), thus preventing detection of the problem in the other branch in the well conditioned (random) cases. Hence, the probability of testing the explicit calls to \texttt{PSCNRM2}/\texttt{PDZNRM2} (and detecting the error) was rather tiny. 

\subsection{From \scalapack forum (\scalapack Archives, October 2011)}\label{SS=DWang}
It should be noted that computational practitioners have already experienced and reported that \texttt{PxGEQPF} sometimes returns badly structured triangular factor. For example, 
in the LAPACK forum\footnote{ 
\texttt{http://icl.cs.utk.edu/lapack-forum/archives/scalapack/msg00254.html}}
David Wang wrote on October 19, 2011.~the following:

\begin{verbatim}
Hello Scalapack,

I have been using the pivoted QR factorisation routine (pzgeqpf) from
scalapack-1.8.0.  I have found that, in some cases, the diagonal
elements of the R matrix are not arranged in order of descending
magnitude as described in the user's guide.  I have attached a test
case for you to verify.  e.g. I get something like:

R(0,0) -0.707107 + 0i
R(1,1) -0.707107 + 0i
R(2,2) 8.16273e-17 + 0i
R(3,3) -0.243173 + 0i
R(4,4) 4.82605e-17 + 0i

This behaviour can be reproduced for the single precision complex case
(although not with the supplied test matrix).  The same test case
works for the real case (pdgeqpf).  e.g.

R(0,0) -0.707107
R(1,1) -0.707107
R(2,2) 0.5
R(3,3) 3.96991e-17
R(4,4) 2.2336e-17

I have not had any problems with the equivalent lapack routines (zgeqp3),
and have not see any mention of this in the errata.

David
\end{verbatim}
\ 

\noindent We could not access the test matrix used in this post, so we cannot identify the real culprit of the bad pivoting.  We have thoroughly tested the new version of the code, with modifications described in \S \ref{SS=Corrections} and \S \ref{SS=PZ-bug}, and it  has passed all tests.
%\noindent Notice that this post stated that the problem might be difficult to reproduce, and that the \lapack function \texttt{ZGEQP3} worked well (because it had been fixed in 2006.).

%% file: 04_conclusion.tex
\section{Concluding remarks}\label{S=Conclusions}
The problem with the stability of the down-dating and its analysis and solution presented in \cite{drmac-bujanovic-LWN-176-2006}, \cite{DrmBuj08}, \cite{SLWN-2010-1} and in this note are both instructive and worrying. It should be worrying that the output of one of the key computational routines of matrix computations can be drastically changed by inserting a seemingly innocuous \texttt{WRITE} statement in the source code, or by changing the topology of the processes, and that such problem had been around undetected in all major software packages from 1965.~until 2006., and that it is still (in 2019.) present in some libraries. Our work on this problem has resolved the issue in \lapack\cite{DrmBuj08, drmac-bujanovic-LWN-176-2006} , \slicot\cite{SLWN-2010-1} , \scalapack (with this note) and other packages that use these libraries as computing engines (such as, e.g., \matlab). Unfortunately, the problem has not been purged from all relevant libraries. The pure programming bug in complex subroutines (\S \ref{SS=PZ-bug}) is simpler in nature, but equally damaging and equally worrying. 

On the other hand, this is an instructive case study for testing software implementations of numerical methods. 
The developments efforts,  testing and tuning of scientific computing software have been mainly focused on speed; \emph{flops} seem to be more appealing feature than numerical robustness and reliability. Too often is a routine backward error analysis, conveniently expressed in matrix norms, considered to be a theoretical certificate for the implementation, which is then tested by checking that the norm of the residual is small. Verifying that the residual is small for many randomly generated matrices is necessary but not necessarily sufficient condition. 
So, for example,  in a followup to the post cited in \S \ref{SS=DWang}, David Wang wrote

\begin{verbatim}
...    Also, the output matrices still multiply to give the original matrix.
Only the pivoting appears to be wrong.
\end{verbatim}

So, both errors, one originating from the first release of \linpack in the 1970's and the other from the first release of \scalapack in the 1990's, have always passed the residual test, and the affected subroutines have been used in many scientific computing packages that have been (presumably) tested as well. The discussion in \S \ref{SS=Bug2-discuss} shows that the often used testing with random matrices may actually fail to test a particular branch of the code due to the fact that the test cases are well conditioned with high probability. This indicates that the testing of scientific computing software is not adequate, that the implementation phase is detached from the numerical analysis of the finite precision execution of the algorithm. Further, using the software debugging tools to analyze the execution of the code in a vicinity of a singularity of our computational task may be misleading.  {In such cases, perhaps the old fashioned source code printouts and colored markers, together with pencil and paper for an analysis, should be reconsidered as a useful debugging tools.}

On the other hand, the discovery of this problem, as described in Remark \ref{RE-discovery}, shows the benefits of conscientiously stress testing the code under the auspices of numerical analysis and perturbation theory. Such an approach follows the principles advocated by Kahan \cite{kahan-BASCD-2008}, \cite{kahan-STANF50-2008}, \cite{kahan-NEEDEBUG-2011},  \cite{kahan-Boulder-2012}.

Of course, high performance is important and all the efforts to improve the run time are well justified. But it is also the responsibility and the duty of the scientific computing community to strive for robustness and reliability of numerical software that is used in applied sciences and engineering, often as a mission critical factor of an engineering design. The difficulty of the task is best described by the following quote from  \cite{kahan-NEEDEBUG-2011}:
\begin{quote}
{\em 
   In conscientiously tested numerical software, the rarity of roundoff-induced anomalies makes them extremely difficult to find by analysis and/or testing. 
   
   Worse, the anomalies can be simultaneously rare, hard to find, and dense in the data.
}
\end{quote}
This too is both instructive and worrying. 
\section{Acknowledgements}
This correction of \texttt{PxGEQPF} has been long overdue. 
The kick-starter for this note and the production of the new \scalapack code for the column pivoted QR factorization was a discussion with Jack Dongarra during a very inspiring meeting 
\emph{Advances in Numerical Linear Algebra: Celebrating the Centenary of the Birth of James H. Wilkinson},\footnote{\texttt{https://nla-group.org/advances-in-numerical-linear-algebra-2019/}}  organized by the Numerical Linear Algebra Group in the Department of Mathematics at the University of Manchester in May 2019. 